\documentstyle[12pt]{article}

\catcode`\@=11

\begin{document}

\baselineskip 24pt
\newcommand{\numero}{YCTP-P3 95}
\newcommand{\titre}{FERMION MASSES IN EXTENDED TECHNICOLOUR}
\newcommand{\auteura}{Nick Evans}
\newcommand{\addressa}{ }
\newcommand{\auteurc}{D.A. Ross }
\newcommand{\beq}{\begin{equation}}
\newcommand{\eeq}{\end{equation}}
\newcommand{\Fn}{\mbox{$F(p^2,\Sigma)$}}

\newcommand{\addressc}{ Physics Department\\ Yale University\\ PO Box
208120\\
New Haven\\ CT 06520\\ USA}
\newcommand{\abstrait}{ Talks presented at Beyond The Standard Model
IV
Conference, Lake Tahoe, Dec 94:
NEW FERMION FAMILIES AND PRECISION ELECTROWEAK DATA:

We highlight a trend in the precision electroweak data towards light
new physics and

argue that some spectra of strongly interacting technifamilies  are
plausibly

compatible  with the data.

FERMION MASSES IN EXTENDED TECHNICOLOUR:

An ETC model with a minimal number of operators responsible for
fermion  masses that

break the global symmetries in the observed manner postdicts  the
light quark masses.

The up down mass inversion may be explained by

the inclusion of a family symmetric ansatz  for the CKM matrix
angles. }

\begin{titlepage}
\hfill \numero
\vspace{.5in}
\begin{center}
{\large{\bf \titre }}
 \bigskip \\by\bigskip\\ \auteura \bigskip \\ \addressc \\

\renewcommand{\thefootnote}{ }
\vspace{.9 in}
{\bf Abstract}
\end{center}
\abstrait
\bigskip \\
\end{titlepage}





 \catcode`\@=11
\long\def\@makefntext#1{
\protect\noindent \hbox to 3.2pt {\hskip-.9pt
$^{{\ninerm\@thefnmark}}$\hfil}#1\hfill}		

\def\@makefnmark{\hbox to 0pt{$^{\@thefnmark}$\hss}}  

\def\ps@myheadings{\let\@mkboth\@gobbletwo
\def\@oddhead{\hbox{}
\rightmark\hfil\ninerm\thepage}
\def\@oddfoot{}\def\@evenhead{\ninerm\thepage\hfil
\leftmark\hbox{}}\def\@evenfoot{}
\def\sectionmark##1{}\def\subsectionmark##1{}}

\setcounter{footnote}{0}
\renewcommand{\thefootnote}{\fnsymbol{footnote}}

\newcounter{sectionc}\newcounter{subsectionc}\newcounter{subsubsection
c}
\renewcommand{\section}[1] {\vspace*{0.6cm}\addtocounter{sectionc}{1}
\setcounter{subsectionc}{0}\setcounter{subsubsectionc}{0}\noindent
	{\normalsize\bf\thesectionc. #1}\par\vspace*{0.4cm}}
\renewcommand{\subsection}[1]
{\vspace*{0.6cm}\addtocounter{subsectionc}{1}
	\setcounter{subsubsectionc}{0}\noindent
	{\normalsize\it\thesectionc.\thesubsectionc.
#1}\par\vspace*{0.4cm}}
\renewcommand{\subsubsection}[1]
{\vspace*{0.6cm}\addtocounter{subsubsectionc}{1}
	\noindent
{\normalsize\rm\thesectionc.\thesubsectionc.\thesubsubsectionc.
	#1}\par\vspace*{0.4cm}}
\newcommand{\nonumsection}[1]
{\vspace*{0.6cm}\noindent{\normalsize\bf #1}
	\par\vspace*{0.4cm}}

\newcounter{appendixc}
\newcounter{subappendixc}[appendixc]
\newcounter{subsubappendixc}[subappendixc]
\renewcommand{\thesubappendixc}{\Alph{appendixc}.\arabic{subappendixc}
}
\renewcommand{\thesubsubappendixc}
	{\Alph{appendixc}.\arabic{subappendixc}.\arabic{subsubappendix
c}}

\renewcommand{\appendix}[1] {\vspace*{0.6cm}
        \refstepcounter{appendixc}
        \setcounter{figure}{0}
        \setcounter{table}{0}
        \setcounter{equation}{0}
        \renewcommand{\thefigure}{\Alph{appendixc}.\arabic{figure}}
        \renewcommand{\thetable}{\Alph{appendixc}.\arabic{table}}
        \renewcommand{\theappendixc}{\Alph{appendixc}}

\renewcommand{\theequation}{\Alph{appendixc}.\arabic{equation}}
        \noindent{\bf Appendix \theappendixc #1}\par\vspace*{0.4cm}}
\newcommand{\subappendix}[1] {\vspace*{0.6cm}
        \refstepcounter{subappendixc}
        \noindent{\bf Appendix \thesubappendixc.
#1}\par\vspace*{0.4cm}}
\newcommand{\subsubappendix}[1] {\vspace*{0.6cm}
        \refstepcounter{subsubappendixc}
        \noindent{\it Appendix \thesubsubappendixc. #1}
	\par\vspace*{0.4cm}}

\def\abstracts#1{{

\centering{\begin{minipage}{12.2truecm}\footnotesize\baselineskip=12pt
\noindent
	\centerline{\footnotesize ABSTRACT}\vspace*{0.3cm}
	\parindent=0pt #1
	\end{minipage}}\par}}

\newcommand{\bibit}{\it}
\newcommand{\bibbf}{\bf}
\renewenvironment{thebibliography}[1]
	{\begin{list}{\arabic{enumi}.}
	{\usecounter{enumi}\setlength{\parsep}{0pt}
\setlength{\leftmargin 1.25cm}{\rightmargin 0pt}
	 \setlength{\itemsep}{0pt} \settowidth
	{\labelwidth}{#1.}\sloppy}}{\end{list}}

\topsep=0in\parsep=0in\itemsep=0in
\parindent=1.5pc

\newcounter{itemlistc}
\newcounter{romanlistc}
\newcounter{alphlistc}
\newcounter{arabiclistc}
\newenvironment{itemlist}
    	{\setcounter{itemlistc}{0}
	 \begin{list}{$\bullet$}
	{\usecounter{itemlistc}
	 \setlength{\parsep}{0pt}
	 \setlength{\itemsep}{0pt}}}{\end{list}}

\newenvironment{romanlist}
	{\setcounter{romanlistc}{0}
	 \begin{list}{$($\roman{romanlistc}$)$}
	{\usecounter{romanlistc}
	 \setlength{\parsep}{0pt}
	 \setlength{\itemsep}{0pt}}}{\end{list}}

\newenvironment{alphlist}
	{\setcounter{alphlistc}{0}
	 \begin{list}{$($\alph{alphlistc}$)$}
	{\usecounter{alphlistc}
	 \setlength{\parsep}{0pt}
	 \setlength{\itemsep}{0pt}}}{\end{list}}

\newenvironment{arabiclist}
	{\setcounter{arabiclistc}{0}
	 \begin{list}{\arabic{arabiclistc}}
	{\usecounter{arabiclistc}
	 \setlength{\parsep}{0pt}
	 \setlength{\itemsep}{0pt}}}{\end{list}}

\newcommand{\fcaption}[1]{
        \refstepcounter{figure}
        \setbox\@tempboxa = \hbox{\footnotesize Fig.~\thefigure. #1}
        \ifdim \wd\@tempboxa > 6in
           {\begin{center}
        \parbox{6in}{\footnotesize\baselineskip=12pt Fig.~\thefigure.
#1}
            \end{center}}
        \else
             {\begin{center}
             {\footnotesize Fig.~\thefigure. #1}
              \end{center}}
        \fi}

\newcommand{\tcaption}[1]{
        \refstepcounter{table}
        \setbox\@tempboxa = \hbox{\footnotesize Table~\thetable. #1}
        \ifdim \wd\@tempboxa > 6in
           {\begin{center}
        \parbox{6in}{\footnotesize\baselineskip=12pt Table~\thetable.
#1}
            \end{center}}
        \else
             {\begin{center}
             {\footnotesize Table~\thetable. #1}
              \end{center}}
        \fi}

\def\@citex[#1]#2{\if@filesw\immediate\write\@auxout
	{\string\citation{#2}}\fi
\def\@citea{}\@cite{\@for\@citeb:=#2\do
	{\@citea\def\@citea{,}\@ifundefined
	{b@\@citeb}{{\bf ?}\@warning
	{Citation `\@citeb' on page \thepage \space undefined}}
	{\csname b@\@citeb\endcsname}}}{#1}}

\newif\if@cghi
\def\cite{\@cghitrue\@ifnextchar [{\@tempswatrue
	\@citex}{\@tempswafalse\@citex[]}}
\def\citelow{\@cghifalse\@ifnextchar [{\@tempswatrue
	\@citex}{\@tempswafalse\@citex[]}}
\def\@cite#1#2{{$\null^{#1}$\if@tempswa\typeout
	{IJCGA warning: optional citation argument
	ignored: `#2'} \fi}}
\newcommand{\citeup}{\cite}

\font\twelvebf=cmbx10 scaled\magstep 1
\font\twelverm=cmr10  scaled\magstep 1
\font\twelveit=cmti10 scaled\magstep 1
\font\elevenbfit=cmbxti10 scaled\magstephalf
\font\elevenbf=cmbx10     scaled\magstephalf
\font\elevenrm=cmr10      scaled\magstephalf
\font\elevenit=cmti10     scaled\magstephalf
\font\bfit=cmbxti10
\font\tenbf=cmbx10
\font\tenrm=cmr10
\font\tenit=cmti10
\font\ninebf=cmbx9
\font\ninerm=cmr9
\font\nineit=cmti9
\font\eightbf=cmbx8
\font\eightrm=cmr8
\font\eightit=cmti8

\textwidth 6.0in
\textheight 8.6in
\pagestyle{empty}
\topmargin -0.25truein
\oddsidemargin 0.30truein
\evensidemargin 0.30truein
\parindent=1.5pc
\baselineskip=15pt



\newcommand{\st}{\scriptstyle}
\newcommand{\sst}{\scriptscriptstyle}
\newcommand{\mco}{\multicolumn}
\newcommand{\epp}{\epsilon^{\prime}}
\newcommand{\vep}{\varepsilon}
\newcommand{\ra}{\rightarrow}
\newcommand{\ppg}{\pi^+\pi^-\gamma}
\newcommand{\vp}{{\bf p}}
\newcommand{\ko}{K^0}
\newcommand{\kb}{\bar{K^0}}
\newcommand{\al}{\alpha}
\newcommand{\ab}{\bar{\alpha}}
\def\be{\begin{equation}}
\def\ee{\end{equation}}
\def\bea{\begin{eqnarray}}
\def\eea{\end{eqnarray}}
\def\CPbar{\hbox{{\rm CP}\hskip-1.80em{/}}}

\input epsf
\newwrite\ffile\global\newcount\figno \global\figno=1
\def\writedefs{\immediate\openout\lfile=labeldefs.tmp
\def\writedef##1{%
\immediate\write\lfile{\string\def\string##1\rightbracket}}}
\def\writestoppt{}\def\writedef#1{}

\def\figin{\epsfcheck\figin}\def\figins{\epsfcheck\figins}
\def\epsfcheck{\ifx\epsfbox\UnDeFiNeD
\message{(NO epsf.tex, FIGURES WILL BE IGNORED)}
\gdef\figin##1{\vskip2in}\gdef\figins##1{\hskip.5in}
instead
\else\message{(FIGURES WILL BE INCLUDED)}%
\gdef\figin##1{##1}\gdef\figins##1{##1}\fi}

\def\figinsert{}
\def\ifig#1#2#3{\xdef#1{fig.~\the\figno}
\writedef{#1\leftbracket fig.\noexpand~\the\figno}%
\figinsert\figin{\centerline{#3}}\medskip\centerline{\vbox
{\baselineskip12pt
\advance\hsize by -1truein\center\footnotesize
{  Fig.~\the\figno.} #2}}
\bigskip\endinsert\global\advance\figno by1}
\def\footnotefont{}\def\endinsert{}

\centerline{\normalsize\bf  NEW FERMION FAMILIES AND PRECISION}
\baselineskip=16pt
\centerline{\normalsize\bf ELECTROWEAK DATA}

\centerline{\footnotesize NICK EVANS}
\baselineskip=13pt
\centerline{\footnotesize\it Physics Department,Yale University,
PO Box 208120,}
\baselineskip=12pt
\centerline{\footnotesize\it New Haven, CT 06520, USA}

\vspace*{0.9cm}
\abstracts{We highlight a trend in the precision electroweak data
towards light
new physics and argue that some spectra of strongly interacting
technifamilies
are plausibly compatible with the data.}

\vspace*{0.6cm}
\normalsize\baselineskip=15pt
\setcounter{footnote}{0}
\renewcommand{\thefootnote}{\alph{footnote}}

In the absence of a theoretical understanding of fermion masses we
can not
exclude the possibility of extra fermion families though the
associated
neutrinos must have masses in excess of 45GeV. Such families   occur
in
strongly interacting models in which electroweak symmetry is broken
by fermion
condensates such as technicolour\cite{technicolour}.The light fermion
masses
are generated by interactions with the fermion condensates mediated
by massive
gauge bosons. If these gauge bosons are EW singlets then there must
be a
condensing fermion with the same quantum numbers as each of the light
fermions
that acquires a mass and hence a strongly interacting fourth family.

Such heavy fermions do not decouple from oblique corrections to the
broken
gauge interactions of the Standard Model\cite{non,Peskin}. We define
the
contribution to the self energy between gauge bosons $X^{\mu}$ and
$Y^{\nu}$ by

\be \int d^4x e^{-iqx} <J^{\mu}_X(x) J^{\nu}_Y(0)> = i \left[
\Pi_{XY}(0) +
q^2\Pi'_{XY}(0) \right] g^{\mu \nu} + q^{\mu} q^{\nu} {\rm terms} \ee

\noindent where we have performed a Taylor expansion (explicitly
assuming that
the new physics is much more massive than the Z mass) and the prime
indicates
differentiation with respect to $q^2$. There are then two observables
in Z pole
measurements, Peskin's S and T\cite{Peskin}, which are combinations
of
$\Pi_{ZZ}(0)$, $\Pi'_{ZZ}(0)$, $\Pi_{ZA}'(0)$, $\Pi_{AA}(0)$ and
$\Pi_{WW}(0)$
in which divergences cancel. In Fig 1a the right hand region
corresponds to the
contributions to S and T from a weakly interacting fourth fermion
family for
all possible mass spectra in excess of 150GeV. The ellipses are the
experimentally viable region of the ST plane\cite{Caravaglios} at one
and two
standard deviations assuming a top mass of 150GeV and a Higgs mass of
300GeV.
Such a scenario is disfavoured by the data.

The experimental lower bound on new fermion masses is, however,
45GeV. If a
member of the extra family has a mass close to this bound then the
Taylor
expansion  will break down. The errors induced may be measured by the
parameters\cite{Burgess}

\begin{eqnarray}
V & = & \Pi_{ZZ}'(0) - \left[ \Pi_{ZZ}(M_Z) - \Pi_{ZZ}(0) \right] /
M_Z^2
\nonumber \\
X & = & \Pi_{ZA}'(0) - \Pi_{ZA}(M_Z)/M_Z^2 \end{eqnarray}

$\left. \right.$ \hspace{-4.8cm} \ifig\prtbdiag{ }
{\epsfxsize7.8truecm\epsfbox{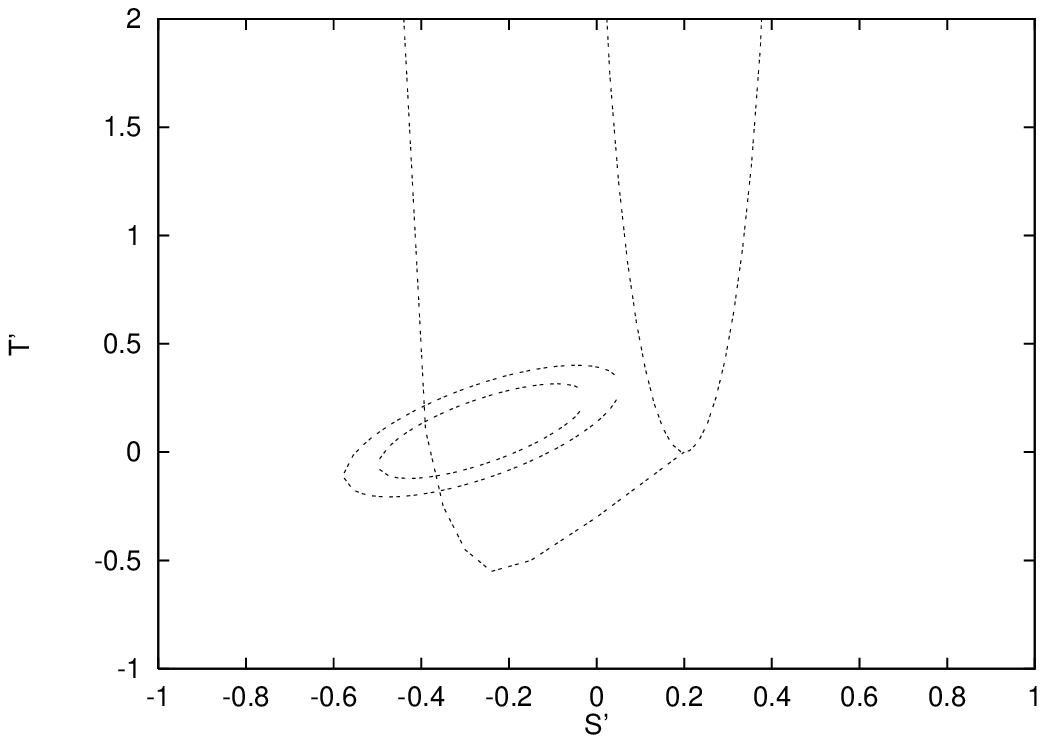}} \vspace{-6.7cm}

$\left. \right.$ \hspace{2.8cm}\ifig\prtbdiag{}
{\epsfxsize7.8truecm\epsfbox{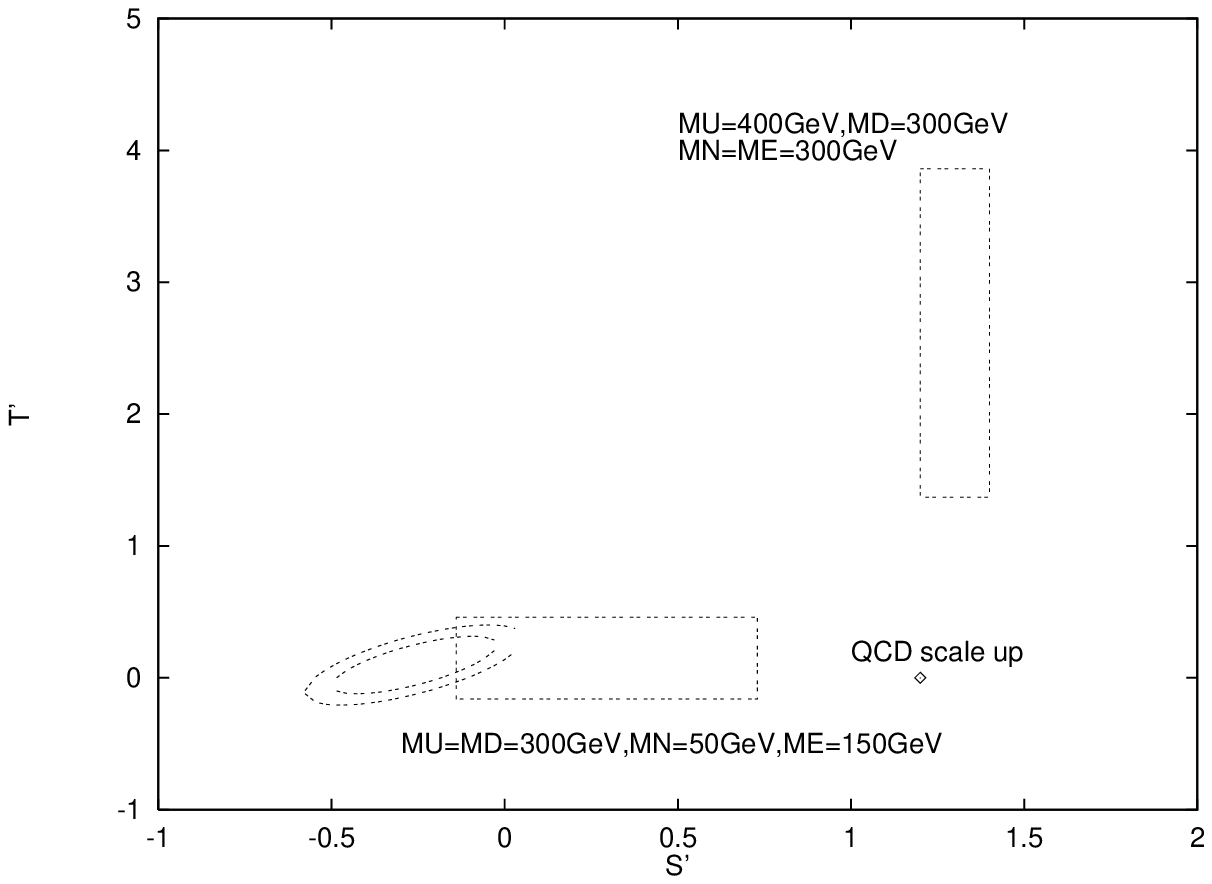}} \vspace{-0.85cm}

\fcaption{a) ST plane showing contribution from heavy and light 4th
family  b)
S'T' plane showing\\ \indent contribution of technifamily spectra}
\vspace{.5cm}

\noindent There are still two observables which may be compared with
the data
but they are now given by S' and T', linear combinations of S,T,V and
X. The
left hand region in Fig 1a shows the contribution to S' and T' from a
fourth
weakly interacting family of fermions with one or more member with a
mass below
150GeV. The data prefers such models with light new physics.

The dynamics of a strongly interacting technifamily below the
confinement scale
may be described by a chiral lagrangian. The $SU(8)_L \otimes
SU(8)_R$
approximate global symmetry of the family is broken to $SU(8)_V$ by
the
formation of condensates and the low energy theory will therefore
have 63
Goldstone like fields

\be {\cal L} = \frac{f_{\pi}^2}{4} tr \left[ \partial^{\mu}\Sigma
\partial_{\mu} \Sigma^{\dagger} \right] + ... \ee

\noindent where $\Sigma = exp[i \pi^a \lambda^a/f_{\pi}]$ with
$\pi^a$ the
Goldstones associated with the generators of SU(8), $\lambda^a$. To
calculate
low energy observables we must match the effective theory to the
underlying
strong dynamics. If the strong interactions are similar to QCD then
we can
match the effective theory to a scaled up version of QCD\cite{Peskin}
and
obtain the contributions from the condensing fields $T = 0$ and  $S
= 0.1/{\rm
doublet}$.
 Such a one family model gives too large a contribution to the S
parameter.
However, QCD is an isospin preserving theory whilst we know that the
SU(8)
vector symmetry is badly broken in the third family masses and hence
presumably
by the electroweak symmetry breaking dynamics. Such symmetry breaking
effects
will very plausibly feed into the technifermion masses as well. We
shall
approximate the strong interactions using Dynamical Perturbation
Theory\cite{DPT}; the technifermions acquire momentum dependent mass
terms,
$\Sigma(k^2)$, consistent with a gap equation analysis and we use the
minimal
technifermion-gauge boson couplings consistent with the chiral Ward
identities.
Estimates of T in this approximation give  $T_{{\rm pert}} <  T  <
2T_{{\rm
pert}}$ \noindent and in the custodial SU(2) limit $S = 0.1/$doublet
in
agreement with the QCD scale up. DPT estimates for S are still work
in progress
but we may bound the solutions by $2 S_{{\rm pert}} < S< S_{{\rm
pert}} + 0.05
$.
Both extremes agree with the QCD scale up but whilst one assumes the
strong
dynamics is blind to isospin breaking the other assumes it is as
sensitive as
the weakly interacting result.

The $SU(8)_L \otimes SU(8)_R$ global symmetry is only an approximate
symmetry
and hence the Goldstone fields will acquire small masses from the
perturbing
interactions. These perturbations are the $SU(3)_C \otimes U(1)_Y$
Standard
Model interactions\cite{gauge} and the extended gauge interactions
responsible
for the third family masses\cite{Casal}. We introduce the former by
gauging the
chiral lagrangian and the latter by including Yukawa interactions
between the
third family fermions and the vev of the $\Sigma$ field. Goldstone
boson masses
then occur at one loop and give masses from the gauge interactions
that are
small relative to the strong interaction scale ($\sim 1 TeV$) and
from the
Yukawa interactions that are small relative to the ETC scale ($\sim
5TeV$). The
lightest Goldstone has a mass very plausibly above the LEP lower
bound.

We may now calculate S and T including both technifermion and pseudo
Goldstone
boson loops for a one family technicolour model\cite{Evans}. We show
the
numerical results for S and T for three strongly interacting SU(3)
technicolour
models in Fig 1b. Spectra with technineutrino masses close to the
current LEP
limit and degenerate techniquarks are plausibly compatible with the
data. It is
reasonable to ask whether this spectrum is compatible with a
technicolour
model of the third family masses. In general each technifermion and
it's third
family counter part may have independent ETC self interactions as
well as the
sideways interaction that feeds the technifermion condensate down to
give the
third family mass. There are  thus three couplings for each flavour
that may be
tuned to give the technifermion mass above and the observed third
family mass
and therefore such a spectrum is not unimaginable. A gap equation
analysis of
the quark sector\cite{Evans2} shows that the techniquark degeneracy
can only be
maintained simultaneously with the large top bottom mass splitting if
the bulk
of the top quark mass is generated by a close to critical top quark
self
interaction.
\newpage

\centerline{\normalsize\bf  FERMION MASSES IN EXTENDED TECHNICOLOUR}

\centerline{\footnotesize NICK EVANS}
\baselineskip=13pt
\centerline{\footnotesize\it Physics Department,Yale University,
PO Box 208120,}
\baselineskip=12pt
\centerline{\footnotesize\it New Haven, CT 06520, USA}

\vspace*{0.9cm}
\abstracts{An ETC model with a minimal number of operators
responsible for
fermion masses that break the global symmetries in the observed
manner
postdicts the light quark masses. The up down mass inversion may be
explained
by the inclusion of a family symmetric ansatz  for the CKM matrix
angles.}

\vspace*{0.6cm}
\normalsize\baselineskip=15pt
\setcounter{footnote}{0}
\renewcommand{\thefootnote}{\alph{footnote}}

Dynamical models in which  electroweak symmetry (EWS) is broken by a
strongly
interacting fermion condensate, such as
technicolour\cite{technicolour}, are
very appealing. These models rely on physics already realized in
nature by QCD
and, since there are no fundamental scalars, there is no hierarchy
problem. The
light fermion masses may be included in the theory by extending the
gauge
sector so that the light fermions may interact with technifermion
condensates
through the exchange of massive ETC gauge bosons ($m_f
=g_{ETC}^2<\bar{T} T> /
M^2_{ETC}$) explaining their masses' suppression relative to the EWS
breaking
scale. Given the almost maximal breaking of the $SU(24)_L \otimes
SU(24)_R$
global symmetry of the light fermions we should not expect to
immediately
understand the extended gauge sector without experimental input.

The large top mass ($m_t \sim 170GeV$) may be such a hint. The
standard
perturbative ETC mass generation breaks down for the top since the
ETC scale
must be of order 1TeV, the scale at which technicolour  becomes
strong. We
conclude that the ETC interactions of the third family may themselves
be
strong. The small deviation of  $\delta \rho$ from zero is also hard
to
reconcile with the large top bottom mass splitting in the standard
ETC picture.
It has been proposed\cite{top} that the  large top mass is the result
of a
strong self interaction. These ETC interactions can not be fine
tuned\cite{chivukula} though since there is an ETC bound state of top
that
becomes a Goldstone when the coupling rises above critical. Below
critical the
bound state's mass is
$ M_{\pi} = (1 -g/g_c) M_{ETC} $.
 Since this light scalar is not observed we must have $g/g_C < 0.9$.

It is interesting to build models\cite{Evans1} of the fermion masses
with
strong ETC interactions. We wish though to avoid specifying the
precise
extended gauge sector for which we have so little data and
concentrate on the
light fermion masses. For this reason we shall represent the heavy
gauge
interactions by 4 Fermi operators. We shall also neglect the neutrino
masses in
the theory since there is clearly something quirky in this sector and
put aside
the CKM mixing angles to begin with.  Let us start by considering the
techi and
third families only. Electroweak symmetry will be broken by a
technifamily
transforming under some $SU(N)_{TC}$ group. The ETC interactions must
then feed
mass to the third family from the technifamily, the quarks' masses
must be
split from the tau and the top from the bottom. We shall introduce
the minimal
number of new ETC operators that break these symmetries; a single
sideways
interaction connecting each of the third family members and their
techni-counter part will give the third family mass;  quark lepton
symmetry
will be broken by a quark self interaction shared by all quarks in
the model;
custodial isospin will be broken  in the quark sector by a top self
interaction. Explicitly the 4 Fermi operators are
\begin{eqnarray}
&& \frac{g_{3}^2}{M_{ETC}^2} {\bar \Psi}_L E_R \bar{\tau}_{R}
\psi_{3L}
\hspace{0.5cm}
\frac{g_{3}^2}{M^2_{ETC}} {\bar Q}_L U_R \bar{t}_{R} q_{3L}
\hspace{0.5cm}
\frac{g_{3}^2}{M^2_{ETC}} {\bar Q}_L D_R \bar{b}_{R} q_{3L}
\nonumber \\
&& \frac{g_{Q}^2}{M^2_{ETC}} {\bar Q}_L U_R \bar{U}_{R} Q_L
\hspace{0.5cm}
\frac{g_{Q}^2}{M^2_{ETC}} {\bar Q}_L D_R \bar{D}_{R} Q_L
\hspace{0.5cm}
\frac{g_{Q}^2}{M^2_{ETC}} {\bar q}_{3L} t_R \bar{t}_{R} q_{3L}
\hspace{0.5cm}
\frac{g_{Q}^2}{M^2_{ETC}} {\bar q}_{3L} b_R \bar{b}_{R} q_{3L}
\nonumber \\
&&\frac{g_{t}^2}{M^2_{ETC}} {\bar q}_{3L} t_R \bar{t}_{R} q_{3L}
\end{eqnarray}
The model has four parameters ($\Lambda_{TC}$, $g_3$, $g_Q$ and
$g_t$) that may
be tuned to give the observed Z, top, bottom and tau masses (any
fewer
parameters would introduce global symmetries that do not exist in the
observed
masses). The technifermion masses are therefore predictions of the
model.

To estimate these masses we must approximate  the non perturbative
dynamics. We
 use the standard maleficia, the gap equation. Although the
uncertainties
associated with the truncation of the Swinger Dyson equations are
large  the
self energies are fixed somewhat by the integral equations for the Z
and third
family masses. We shall indicate the size of errors by using extreme
functions
for the technicolour running coupling in the non perturbative regime
as an
example. Our ansatzes for the coupling are that at high momentum it
runs
perturbatively to either the critical coupling or three times the
critical
coupling at $\Lambda_{TC}$ and then below $\Lambda_{TC}$ the running
is  cut
off. Solving the gap equations  gives for an $SU(3)_{TC}$ group the
technifermion masses
\be 300 GeV < M_Q < 400GeV,  \hspace{0.25cm} \Delta M_Q = 20 \pm
15GeV,
\hspace{0.5cm} 200GeV < M_E < 300GeV \ee
\indent This model with a minimal number of operators may be simply
extended to
the first and second families by the inclusion of a single extra
sideways
coupling for each family connecting its members to the associated
heavier
fermions. For example for the second family \vspace{-0.2cm}

\begin{eqnarray}
&&\frac{g_{2}^2}{M_{ETC}^2} {\bar \Psi}_L E_R \bar{\mu}_{R} \psi_L
\hspace{0.5cm}
\frac{g_{2}^2}{M^2_{ETC}} {\bar Q}_L U_R \bar{c}_{R} q_L
\hspace{0.5cm}
\frac{g_{2}^2}{M^2_{ETC}} {\bar Q}_L D_R \bar{s}_{R} q_L \nonumber \\
&&
\frac{g_{2}^2}{M^2_{ETC}} {\bar \psi}_{3L} \tau_R \bar{\mu}_{R}
\psi_{2L}
\hspace{0.5cm}
\frac{g_{2}^2}{M^2_{ETC}} {\bar q}_{3L} t_R \bar{c}_{R} q_{2L}
\hspace{0.5cm}
\frac{g_{2}^2}{M^2_{ETC}} {\bar q}_{3L} b_R \bar{s}_{R} q_{2L}
\end{eqnarray}
 $g_2$ is fixed by requiring that the correct muon mass is obtained
(and
similarly for the first family by requiring the correct electron
mass). The
model then postdicts the light quark masses. We obtain
\begin{eqnarray}
& m_c = 1.5 \pm 0.8 GeV,  & m_s = 0.32\pm 0.02GeV, \nonumber \\
  & m_u = 6.6\pm3.7MeV, & m_d =1.5 \pm 0.2 MeV \end{eqnarray}
These estimates are in surprisingly good agreement with the
experimental values
although the up down mass inversion is not reproduced. Of course
there are many
models currently in the literature postdicting fermion masses at
least some of
which must be numerological coincidences but hopefully this model
demonstrates
that ETC is potentially a predicitve model of fermion masses.

Finally let us return to the CKM mixing angles  that we have so far
neglected.
The most plausible explanation of the three family scales is to
associate them
with three separate ETC breaking scales of some $SU(N+3) $ ETC group.
However,
if a single gauge eigenstate is picked out by the breaking at each
scale then,
since it is the gauge interactions that generate the fermion masses,
there will
be definite mass eigenstates. In such a model the off diagonal mass
terms must
be generated by additional dynamics. In the spirit of the above
analysis let us
propose a simple ansatz for this additional mass generation that has
an
$SU(3)_L \otimes SU(3)_R$ family symmetry and is flavour blind. The
extra
contributions to the mass matrices will therefore take the form of
some new
scale ($\nu$) times an SU(3) matrix.This ansatz may be coerced to fit
the
observed CKM data\cite{Evans2} but of particular interest is its
ability to
generate the up down mass inversion.  Concentrating on the Cabbibo
sector and
choosing $\sigma_2$ as the SU(2) subgroup of SU(3) we have
\be
M_U   =   \left( \begin{array}{cc} 1.5 & 0 \\ 0 & 0.005 \end{array}
\right) +
\nu  \left( \begin{array}{cc} 0 & 1 \\ 1& 0  \end{array} \right),
\hspace{0.3cm}
 M_D   =   \left( \begin{array}{cc} 0.3 & 0 \\ 0 & 0.002 \end{array}
\right) +
\nu  \left( \begin{array}{cc} 0 & 1 \\ 1& 0  \end{array} \right)
\ee
 where in the first, ETC generated, matrices the up down masses are
not
inverted. Diagonalizing and searching for a value of $\nu$ compatible
with the
Cabbibo angle we find
\be   \begin{array}{cc} m_c \sim 1.5GeV, & m_u \sim 0.003GeV \\  m_s
\sim
0.3GeV, & m_d \sim 0.013GeV \end{array} \hspace{0.5cm} | C | = \left(
\begin{array}{cc} 0.975 & 0.22 \\ 0.22 & 0.975 \end{array} \right)
\ee

\noindent Simply models of this form are however flawed since the
effective
potential receives contributions from loops of light fermions which,
if they
have two separate contributions to their mass generation, give rise
to terms of
the form $-tr(M_1 M_2)$ which prefers to simultaneously diagonalize
$M_1$ and
$M_2$. More complicated models may though be able to stabilize
realistic vacua.

\end{document}